\documentclass[11pt,a4paper]{article}

\usepackage{amsmath,amssymb,amsfonts,amsthm,bbm,bm,color,array} 
\usepackage[T1]{fontenc}
\usepackage{lmodern}
\usepackage[utf8]{inputenc} 
\usepackage[protrusion=true,expansion=true]{microtype}
\usepackage{textcomp} 

\usepackage{authblk}

\usepackage[left=2.58cm, right=2.58cm, top=2.54cm]{geometry}
\usepackage[pdftex,bookmarks,colorlinks,breaklinks]{hyperref}  
\usepackage{xcolor}
\definecolor{dullmagenta}{rgb}{0.4,0,0.4}  
\definecolor{darkblue}{rgb}{0,0,0.4}
\hypersetup{linkcolor=red,citecolor=blue,filecolor=dullmagenta,urlcolor=darkblue} 

\usepackage{pdfsync}

\newcommand{\ints}{\mathbb Z}
\newcommand{\HH}{\mathcal{H}}
\newcommand{\WH}{\mathrm{WH}}
\newcommand{\SL}{\mathrm{SL}}
\newcommand{\1}{\mathbbm{1}}
\newcommand{\ket}[1]{|{#1}\rangle}
\newcommand{\bra}[1]{\langle #1 |}
\newcommand{\braket}[2]{\langle #1|#2\rangle}
\newcommand{\ketbra}[2]{|#1\rangle\langle #2|}
\newcommand{\tr}{\operatorname{tr}}
\newcommand{\Rk}{\operatorname{rank}}
\newcommand{\spec}{\operatorname{spec}}
\newcommand{\md}{\text{ mod }}
\newcommand{\XXX}{\mathbb{X}}
\newcommand{\ZZZ}{\mathbb{Z}}
\newcommand{\FFF}{\mathbb{F}}
\newcommand{\UUU}{\mathbb{U}}
\newcommand{\VVV}{\mathbb{V}}
\newcommand{\WWW}{\mathbb{W}}
\newcommand{\III}{\mathbb{I}}

\title{Aligned SICs and embedded tight frames in even dimensions}
\author{Ole Andersson\footnote{ole.andersson@fysik.su.se} }
\author{Irina Dumitru\footnote{irina.dumitru@fysik.su.se} }
\affil{Department of Physics, Stockholm University, 106 91 Stockholm, Sweden}
\date{}
\begin{document}
\maketitle

\begin{abstract}
\noindent 
Alignment is a geometric relation between pairs of Weyl-Heisenberg SICs, 
one in dimension $d$ and another in dimension $d(d-2)$, 
manifesting a well-founded conjecture about a number-theoretical 
connection between the SICs. In this paper, we prove that if $d$ is even, 
the SIC in dimension $d(d-2)$ of an aligned pair can be partitioned into 
$(d-2)^2$ tight $d^2$-frames of rank $d(d-1)/2$ and, alternatively, into 
$d^2$ tight $(d-2)^2$-frames of rank $(d-1)(d-2)/2$. The corresponding 
result for odd $d$ is already known, but the proof for odd $d$ relies on 
results which are not available for even $d$.
We develop methods that allow us to overcome this issue. 
In addition, we provide a relatively 
detailed study of parity operators in the Clifford group, emphasizing differences in the theory 
of parity operators in even and odd dimensions and discussing consequences 
due to such differences. In a final section, we study implications of 
alignment for the symmetry of the SIC.
\end{abstract}

\section{Introduction}
An informationally complete POVM is one that can be used 
to reconstruct any quantum state, pure or mixed. 
Since an $n$-dimensional state is given by an $n\times n$ unit-trace Hermitian matrix, and, hence, by $n^2-1$ real parameters, a minimal
informationally complete POVM has to consist of $n^2$ unit rank elements, giving $n^2-1$ 
independent measurement results. This paper deals with such POVMs.
Specifically, it deals with so-called symmetric informationally complete POVMs \cite{ReB-KScCa2004} (SIC-POVMs, or SICs, for short). 
SICs are exceptional among informationally complete POVMs in the sense that the information overlap of the measurement 
results is minimal, making them optimal candidates for state tomography \cite{scott2006}. These remarkable tomographic properties
reflect that the SIC elements constitute an equiangular tight frame of 
maximally many vectors.

Whether SICs exist in all dimensions is still an open question. 
In his doctoral thesis \cite{Za2011}, G.\ Zauner conjectured that in all finite dimensions 
at least one SIC exists that is covariant under the discrete Weyl-Heisenberg group, 
and he further conjectured that at least one such SIC has an order $3$ unitary symmetry. 
These conjectures have been guiding the search for SIC-POVMs ever since. 
As a result of such searches, we are now confident that we know all Weyl-Heisenberg covariant SICs in Hilbert spaces up to 
dimension $50$ \cite{ScGr2010}, and, interestingly, all of them have the symmetry conjectured by Zauner. 
Furthermore, at least one SIC has been found in each dimension up to $181$ \cite{GrScprivate}, and there are several known SICs in dimensions above that, with the highest dimension being $2208$ \cite{GrScprivate}. 

In this paper, we are interested in properties of Weyl-Heisenberg SIC-POVMs.
In particular, 
we are interested in properties of what we call \emph{aligned} SICs in composite 
dimensions of the form $d(d-2)$. 
Alignment is a geometric relation between a SIC in dimension $d(d-2)$ and a SIC in the corresponding 
dimension $d$ which manifests a conjectured number-theoretical connection 
between SICs in such dimensions \cite{ApFlMcCYa2016}.
The presence of alignment was discovered numerically \cite{ApBeDuFl2017}
by looking at all SICs known at the time in dimensions $d$ 
and $d(d-2)$, the highest value of $d$ being $15$. 
For each SIC in dimension $d$, a SIC in dimension $d(d-2)$ was found to which it is aligned.
In the meantime, the observation of this relation guided the search for a SIC in dimension $323=19(19-2)$ \cite{GrSc2017}. All known aligned SICs in 
composite dimension $d(d-2)$ have also been observed to exhibit a remarkable 
geometric property of their own, namely the embedding of lower-dimensional equiangular 
tight frames \cite{ApBeDuFl2017, Wa2017, ApBeFlGo2019}. 

In dimensions being the product of two relatively prime factors,
each representation of the Weyl-Heisenberg group splits 
into a tensor product representation.
This result was proven in \cite{Gross} using the Chinese Remainder Theorem and 
application of this result 
is, nowadays, referred to as \emph{Chinese remaindering}. 
Chinese remaindering can be applied in odd dimensions of the form we are interested in, since for odd $d$ the factors $d$ and $d-2$ are relatively prime,
and has indeed been used to prove the existence of embedded tight frames in the SIC in the larger dimension of an aligned pair \cite{ApBeDuFl2017}.
However, for even $d$, Chinese remaindering cannot be applied, at least not immediately.
In the current paper, we use special properties of 
representations of the Weyl-Heisenberg group in dimensions divisible by $4$
to overcome this issue (and thereby 
lay out an approach for the treatment of more general composite dimensions whose factors have $2$ 
as the greatest common divisor), and we extend the results in \cite{ApBeDuFl2017} 
to even dimensions of the form $d(d-2)$.
 
Parity operators in the Clifford group play a role in our treatment of aligned SICs, and they too show different behaviors in even and odd dimensions. The differences are similar to those that give rise to a uniqueness issue in the extension of the Wigner function to discrete spaces: The Wigner function can be defined using parity operators \cite{Ro1977}, which allows for an extension to discrete spaces. The extension is canonical in the odd-dimensional case \cite{Gr2006}, but it is not so in the even-dimensional case \cite{ChMuSi2010,ViChChFoKl2018}.

The paper is structured as follows. Section \ref{sec2} deals with the theory of SIC-POVMs and equiangular tight frames and introduces the notion of alignment.
In Section \ref{sec3} we use the apparatus of Chinese remaindering to prove the existence of equiangular
tight frames embedded in aligned SICs. Part of this section is dedicated to a discussion of parity operators. Section \ref{sec4} explores the consequences of alignment for the symmetry of SICs.

\section{Equiangular tight frames and aligned SICs}\label{sec2}
An equiangular tight $m$-frame in an $n$-dimensional Hilbert space is a 
set of unit-length vectors $\ket{\psi_0},\ket{\psi_1},\dots,\ket{\psi_{m-1}}$ 
which satisfies the two conditions
\begin{align}
	&|\braket{\psi_a}{\psi_b}|^2
	=\frac{m-n}{n(m-1)}\,\text{ if }\, a\ne b, \label{equiangular} \\ 
	&\frac{n}{m}\sum_{a=0}^{m-1}\ketbra{\psi_a}{\psi_a}
	=\1 \label{tight}.
\end{align}
That the common angle between any two vectors in the frame has to be the one 
specified in \eqref{equiangular}
follows from the assumption that the frame is normalized and the 
tightness condition \eqref{tight}.
Furthermore, one can show that such a frame can contain neither less than 
$n$ nor more than $n^2$ vectors, see \cite{BeFi2003}.
In the extremal case $m=n$, an equiangular tight $m$-frame 
is the same thing as an orthonormal basis, and 
if $m=n^2$, an equiangular tight $m$-frame is a SIC.
The acronym SIC is a short version of the longer SIC-POVM 
which stands for 
``Symmetric Informationally Complete Positive-Operator Valued Measure''.
As was mentioned in the introduction, such measures have exceptional tomographic properties.
Here, however, we will only be concerned with their geometric characteristics.
For the reader's convenience we repeat the defining conditions satisfied by a SIC:
\begin{align}
	&|\braket{\psi_a}{\psi_b}|^2=\frac{1}{n+1} \text{ if $a\ne b$ },   
	&\frac{1}{n}\sum_{a=0}^{n^2-1}\ketbra{\psi_a}{\psi_a}=\1 \label{SIC tight}.
\end{align}

\subsection{Weyl-Heisenberg SICs and alignment}
Zauner formulated a very strong conjecture in his thesis \cite{Za2011}, namely 
that in every dimension a SIC exists which is an orbit 
under a unitary representation of the discrete Weyl-Heisenberg group. 
He also conjectured that in every dimension a SIC fiducial vector can 
be chosen among the eigenvectors of an operator of order $3$ in the Clifford group,
nowadays referred to as a ``Zauner operator''. 
A SIC fiducial vector is a unit length vector which generates a SIC when the unitaries 
in the Weyl-Heisenberg group displace it, and the Clifford group is the normalizer 
of the Weyl-Heisenberg group, see Section \ref{clifford}.
Almost all known examples of SICs are generated by irreducible representations of the Weyl-Heisenberg group \cite{ScGr2010, FuHoSt2017}, and in this paper we will only consider such SICs. 
We call them Weyl-Heisenberg SICs or WH-SICs for short.

\subsubsection{The Weyl-Heisenberg group}\label{WH group section}
The discrete Weyl-Heisenberg group $\WH(n)$ has three generators $\omega$, $X$, and $Z$. The generators have order $n$, $\omega$
commutes with all the group elements, and the other two generators satisfy the commutation relation $ZX=\omega XZ$. 

Let $(\omega_n, X_n, Z_n)$ be an irreducible unitary representation of $\WH(n)$ on 
an $n$-dimensional Hilbert space (i.e., $\omega_n$, $X_n$, and $Z_n$ are the 
unitary operators corresponding to $\omega$, $X$, and $Z$). Then, by a 
theorem of Weyl \cite[Ch.\ IV, \S 15]{We1930}, $\omega_n$ is a multiple 
of the identity operator, and $X_n$ and $Z_n$ are represented by generalized Pauli matrices relative to an orthonormal basis $\{\ket{u}: u\in\ints_n\}$:
\begin{align}\label{standard}
	&X_n=\sum_{u=0}^{n-1}\ketbra{u+1}{u},
	&Z_n=\sum_{u=0}^{n-1}\omega_n^{u}\ketbra{u}{u}.
\end{align}
The multiplier of the identity in $\omega_n$ (which we also denote by $\omega_n$) can be any primitive $n$th root of unity. 
In this paper, however, we will only consider representations of $\WH(n)$ in which $\omega_n=e^{2\pi i/n}$.

\subsubsection{Displacement operators}
It is convenient for many purposes, including our own, to define so-called 
displacement operators. 
We thus set $\tau_n=-e^{\pi i/n}$ and, for any pair of 
integers $a$ and $b$, define
\begin{equation}\label{displacement operator}
	D_{a,b}^{(n)}=\tau_n^{ab} X_n^aZ_n^b.
\end{equation}
(The superscript $``(n)"$ is to indicate that the displacement operator acts on an 
$n$-dimensional Hilbert space.)
In odd dimensions $\tau_n$ is a power of $\omega_n$. Hence the displacement operators all belong to and generate the representation of
the Weyl-Heisenberg group.
In even dimensions, however, this is not the case, and the group generated by the displacement operators is larger than the representation of the Weyl-Heisenberg group.
The `double-dimensional' order of $\tau_n$ complicates matters. Still, there are reasons, see \cite{Ap2005}, for defining the displacement operators as in \eqref{displacement operator} 
in all dimensions. In any case, the displacement operators generate the same SIC as the Weyl-Heisenberg group when fed with the same SIC fiducial vector.

A straightforward calculation shows that
\begin{equation}\label{merging rule}
	D_{a,b}^{(n)}D_{k,l}^{(n)}=\tau_n^{bk-al} D_{a+k,b+l}^{(n)}.
\end{equation}
From this follows that the Hermitian conjugate of $D_{a,b}^{(n)}$  is $D_{-a,-b}^{(n)}$
and that the displacement operators satisfy the commutation rule
\begin{equation}\label{commutation rule}
	D_{a,b}^{(n)}D_{k,l}^{(n)}=\omega_n^{bk-al}D_{k,l}^{(n)}D_{a,b}^{(n)}.
\end{equation}
The displacement operators also satisfy the translation properties
\begin{align}
	&D_{a+n,b}^{(n)} = (-1)^{(n+1)b}D_{a,b}^{(n)}, 
	&&D_{a,b+n}^{(n)} = (-1)^{(n+1)a}D_{a,b}^{(n)}. \label{disp trans prop}
\end{align}
Thus, they are periodic in the indices
if $n$ is odd, while they are periodic or anti-periodic depending on the parity 
of the index being translated if $n$ is even.

We will frequently use the fact that the displacement operators
(or their Hermitian conjugates) corresponding to indices $0\leq a,b\leq n-1$
form an orthogonal operator basis. The inner product of two displacement operators in the basis is
$\tr\big(D_{-a,-b}^{(n)} D_{k,l}^{(n)}\big)=n\delta_{ak}\delta_{bl}$
and, hence, any operator $A$ can be expanded as
\begin{equation}
	A = \frac{1}{n}\sum_{a,b=0}^{n-1}\tr(D_{-a,-b}^{(n)} A) D_{a,b}^{(n)}
		= \frac{1}{n}\sum_{a,b=0}^{n-1}\tr(D_{a,b}^{(n)} A) D_{-a,-b}^{(n)}.
\end{equation}
This is \emph{the expansion of $A$ in the displacement operator basis}.

\subsubsection{The Clifford group}\label{clifford}
The Clifford group is the normalizer of the Weyl-Heisenberg group in the unitary group. 
In other words, the Clifford group consists of those unitary operators $V$ which are 
such that $VX_nV^\dagger$ and $VZ_nV^\dagger$ belong to the representation of the Weyl-Heisenberg group. 
This definition also determines the Clifford group as an abstract group: By the theorem of Weyl referred to in Section \ref{WH group section}, 
any two irreducible representations of
the Weyl-Heisenberg group (which assign the same value to $\omega$) are canonically unitarily invariant. 
Hence, so are the Clifford groups associated with the different representations. We refer to \cite{Ap2005} for an extensive account
of the relation between the Clifford group and SICs.

Let $\bar n=n$ if $n$ is odd and $\bar n=2n$ if $n$ is even.
The symplectic group $\SL(2,\ints_{\bar n})$, i.e., the group of $2 \times 2$ matrices with entries in the ring of integers modulo $\bar n$ and determinant $1$,
admits a projective representation $F\to V_F$ in the Clifford group, see \cite{Ap2005}.
If 
\begin{equation}
	F=\begin{pmatrix} \alpha & \beta \\ \gamma & \delta \end{pmatrix}
\end{equation}
is a symplectic matrix for which $\beta$ is invertible modulo $\bar n$, then, in the basis relative to which
$X_n$ and $Z_n$ are represented by generalized Pauli matrices \eqref{standard},
\begin{equation}\label{the rep}
	V_F = \frac{1}{\sqrt{n}} \sum_{u,v=0}^{n-1} \tau_n^{\beta^{-1}(\alpha v^2-2uv+\delta u^2)} \ketbra{u}{v}.
\end{equation}
We use the language in \cite{Ap2005} and call symplectic matrices with $\beta$ invertible modulo $\bar n$ \textit{prime}. For non-prime $F$ one can always find 
prime symplectic 
matrices $F_1$ and $F_2$ such that $F=F_1F_2$, see \cite{Ap2005}. We then define
\begin{equation}\label{tut}
	V_F = V_{F_1}V_{F_2}.
\end{equation}
The definition \eqref{tut} together with \eqref{the rep} determines $V_F$ up to a phase,
meaning that different prime decompositions of $F$ give rise to operators $V_F$ which may differ by a phase factor. This indeterminacy is what is meant by the representation being `projective'. Henceforth, we refer to unitary operators of the form $V_F$ as symplectic unitaries. 
The symplectic unitaries satisfy the important identity
\begin{equation}\label{prop of F}
	V_F D_{a,b}^{(n)}  V_F^\dagger = D_{F(a,b)}^{(n)}.
\end{equation}
The indices of the displacement operator in the right-hand side are the entries of the matrix obtained by applying $F$ to $(a,b)^T$. 

\subsection{Alignment}
Let $\ket{\psi_{a,b}}$ be the vector obtained by applying $D_{a,b}^{(n)}$ 
to a SIC fiducial vector $\ket{\psi_{0,0}}$. 
Unless $a=b=0\md n$, the magnitude of the overlap 
between $\ket{\psi_{a,b}}$ and $\ket{\psi_{0,0}}$ is $1/\sqrt{n+1}$.
We define the overlap phases for a WH-SIC in dimension $n$ by
\begin{equation}\label{overlap phases}
	e^{i\theta_{a,b}^{(n)}} = 
	\begin{cases}
		1 & \text{if $a=b=0\md n$,}\\
		\sqrt{n+1}\braket{\psi_{0,0}}{\psi_{a,b}} & \text{otherwise.}
	\end{cases}
\end{equation}

Alignment is a geometric relation between WH-SICs in dimensions $d$ and $n=d(d-2)$
which manifests a conjectured number-theoretical connection between the overlap phases 
of WH-SICs in dimensions $d$ and $d(d-2)$:
We say that a WH-SIC in dimensions $d$ is aligned with a WH-SIC in dimension $n=d(d-2)$ 
if there exist choices of fiducial vectors for 
these such that if $a\ne 0\md (d-2)$ or $b\ne 0\md (d-2)$, then
\begin{equation}\label{alignment condition1}
	e^{i\theta_{da,db}^{(n)}}=
		\begin{cases} 
			1 &\text{if $d$ is odd,} \\ 
			-(-1)^{(a+1)(b+1)} &\text{if $d$ is even,}
		\end{cases}
\end{equation}
and if $a\ne 0\md d$ or $b\ne 0\md d$, then
\begin{equation}\label{alignment condition2}
	e^{i\theta_{(d-2)a,(d-2)b}^{(n)}}=
		\begin{cases} 
			-e^{2i\theta_{\alpha a+\beta b,\gamma a+\delta b}^{(d)}} &\text{if $d$ is odd}, \\ 
			(-1)^{(a+1)(b+1)}e^{2i\theta_{\alpha a+\beta b,\gamma a+\delta b}^{(d)}} &\text{if $d$ is even},
		\end{cases}
\end{equation}
where $\alpha$, $\beta$, $\gamma$, and $\delta$ are integers modulo 
$d$ such that $\alpha\delta-\beta\gamma=\pm 1$.
G.\ McConnell was the first to observe these relations for the phases \cite{Mcunpub}.  The concept of alignment was introduced in \cite{ApBeDuFl2017},
and, supported by extensive numerical and analytical evidence, the authors conjectured that aligned pairs of SICs exist for all 
values of $d$.
It was also proven in \cite{ApBeDuFl2017} that if $d$ is odd, any SIC in dimension $d(d-2)$ which satisfies 
\eqref{alignment condition1} 
can be partitioned into $(d-2)^2$ equiangular tight $d^2$-frames of rank
$d(d-1)/2$, or, alternatively, 
into $d^2$ equiangular tight $(d-2)^2$-frames of rank $(d-1)(d-2)/2$.
Below we prove that the same is true if $d$ is even.

Whether one of the conditions \eqref{alignment condition1}
 and \eqref{alignment condition2} follows from the other is not known.
But no SIC is known which satisfies only one of the conditions.
The results in this paper, however, only rely on \eqref{alignment condition1} being
fulfilled. When we use the expression ``aligned SIC'' we refer to the higher-dimensional member of an aligned pair. 

\subsection{Unitary equivalence}\label{invariance}
Alignment is a property shared among unitarily equivalent WH-SICs. 
Therefore, when examining those intrinsic properties of WH-SICs which are consequences of alignment, one may first apply any suitable unitary to the vectors of the SIC and then proceed with the study.
The theorem of Weyl referred to in Section \ref{WH group section} allows one to do this at the level of representations. For according to that theorem, two irreducible $n$-dimensional representations of $\WH(n)$ which assign the same multiple of the unit operator to $\omega$ are unitarily equivalent. 
We will use this freedom to rotate the representation when convenient.

\section{Equiangular tight frames in aligned SICs}\label{sec3}
Suppose that $\{\ket{\psi_{a,b}}\}$ is an aligned WH-SIC
in dimension $n=d(d-2)$. We prove that if $n$ is even,
the $d^2$-frame
\begin{equation}\label{frame1}
	\{ \ket{\psi_{(d-2)a,(d-2)b}}:a,b=0\dots d-1 \}
\end{equation}
spans and is tight in a $d(d-1)/2$-dimensional space, and the 
$(d-2)^2$-frame
\begin{equation}\label{frame2}
	\{ \ket{\psi_{da,db}}:a,b=0\dots d-3 \}
\end{equation}
spans and is tight in a $(d-1)(d-2)/2$-dimensional space.
By shifting the frame in \eqref{frame1}, respectively \eqref{frame2}, by appropriate displacement operators the SIC gets partitioned into $(d - 2)^2$ equiangular
tight $d^2$-frames, respectively into $d^2$ equiangular tight $(d - 2)^2$-frames.
The corresponding result for odd $n$ was proven in \cite{ApBeDuFl2017}.
Notice that, since 
the equiangularity condition \eqref{equiangular} is automatically satisfied, 
it suffices to prove that 
\begin{align}
	&\Pi_1=\frac{d-1}{2d}\sum_{a,b=0}^{d-1} 
		\ketbra{\psi_{(d-2)a,(d-2)b}}{\psi_{(d-2)a,(d-2)b}}, \label{first projection}\\
	&\Pi_2=\frac{d-1}{2(d-2)}\sum_{a,b=0}^{d-3} \ketbra{\psi_{da,db}}{\psi_{da,db}}, \label{second projection}
\end{align}
are projection operators of rank $d(d-1)/2$ and $(d-1)(d-2)/2$, respectively. 

\subsection{Block-diagonal splitting}\label{block-diagonal splitting}
When $n$ is even, $d$ and $(d-2)$ also have to be even.
We write $d=2n_1$ and $(d-2)=2n_2$. 
The integers $n_1$ and $n_2$ are relatively prime, being consecutive 
integers. In Appendix \ref{unusual} it is shown that, due to this fact,
the Hilbert space can
be decomposed into four 
$(n_1n_2)$-dimensional subspaces, and that there are irreducible representations of 
$\WH(n_1n_2)$ on these subspaces such that the displacement 
operators with even indices are block-diagonal:
\begin{equation}\label{dispdiag}
	D^{(n)}_{2a,2b}
	=(-1)^{ab}
	\begin{pmatrix}
			  D_{a,b}^{(n_1n_2)} & & & \\
			    & \omega_{2n_1n_2}^a D_{a,b}^{(n_1n_2)} & & \\ 
		        & & \omega_{2n_1n_2}^b D_{a,b}^{(n_1n_2)} & \\ 
				& & & \omega_{2n_1n_2}^{a+b} D_{a,b}^{(n_1n_2)}\\
	\end{pmatrix}.
\end{equation}
Furthermore, Chinese remaindering, see Appendix \ref{CRT},
introduces a tensor product in each subspace
which splits it into an $n_1$-dimensional factor and an $n_2$-dimensional factor.
The subspace displacement operators then split according to

\begin{equation}
	D_{a,b}^{(n_1n_2)} = D_{a,\kappa_2b}^{(n_1)}\otimes D_{a,\kappa_1b}^{(n_2)}.
\end{equation}
The integers $\kappa_1$ and $\kappa_2$ are the multiplicative inverses of 
$n_1$ and $n_2$ modulo $\bar n_2$ and $\bar n_1$, respectively. (See Appendix \ref{CRT}.)
We have in particular that
\begin{align}
	&(-1)^{n_1^2ab} D_{n_1a,n_1b}^{(n_1n_2)} 
		= D_{n_1a,\kappa_2n_1b}^{(n_1)}\otimes (-1)^{n_1^2ab} D_{n_1a,\kappa_1n_1b}^{(n_2)}
		= \1_{n_1} \otimes D_{a,b}^{(n_2)}, \\
	&(-1)^{n_2^2ab} D_{n_2a,n_2b}^{(n_1n_2)} 
		= (-1)^{n_2^2ab} D_{n_2a,\kappa_2n_2b}^{(n_1)}\otimes  D_{n_2a,\kappa_1n_2b}^{(n_2)}
		=  D_{-a,b}^{(n_1)} \otimes \1_{n_2} .
\end{align}
These are critical observations for what we intend to show.
The rightmost identities, which hold factor-by-factor, follow from straightforward calculations.
Since
\begin{align}
	&\omega_{2n_1n_2}^{n_1a}= \omega_{2n_2}^{a}, 
	&&\omega_{2n_1n_2}^{n_1b}= \omega_{2n_2}^{b},
	&&\omega_{2n_1n_2}^{n_2a}= \omega_{2n_1}^{a}, 
	&&\omega_{2n_1n_2}^{n_2b}= \omega_{2n_1}^{b},
\end{align}
we have that
\begin{equation}
	D^{(n)}_{da,db}
	= 	\begin{pmatrix} 
		 \1_{n_1} \otimes D_{a,b}^{(n_2)} & & & \\
		 & \1_{n_1} \otimes \omega_{2n_2}^a  D_{a,b}^{(n_2)} & & \\
		 & & \1_{n_1} \otimes \omega_{2n_2}^b  D_{a,b}^{(n_2)} & \\
		 & & & \1_{n_1} \otimes \omega_{2n_2}^{a+b}  D_{a,b}^{(n_2)} 
	  	\end{pmatrix} \label{block1}
\end{equation}
and
\begin{equation}
	D^{(n)}_{(d-2)a,(d-2)b}
 	= \begin{pmatrix}
		D_{-a,b}^{(n_1)} \otimes \1_{n_2}& & & \\
		 & \hspace{-3pt}\omega_{2n_1}^a  D_{-a,b}^{(n_1)}\otimes \1_{n_2} & & \\
		 & & \hspace{-3pt}\omega_{2n_1}^b  D_{-a,b}^{(n_1)}\otimes \1_{n_2} & \\
		 & & & \hspace{-3pt}\omega_{2n_1}^{a+b}  D_{-a,b}^{(n_1)}\otimes \1_{n_2} 
	  \end{pmatrix}. \label{block2}
\end{equation}

\subsubsection{Block diagonal structure of $\Pi_1$ and $\Pi_2$.}
We can now use the decompositions \eqref{block1} and \eqref{block2} to show that 
$\Pi_1$ and $\Pi_2$ are also block-diagonal, and that the blocks have a particular structure.

The expansions of $\Pi_1$ and $\Pi_2$ in the displacement operator basis read
\begin{align}
	&\Pi_1 = \frac{d(d-1)}{2n} \sum_{a,b=0}^{d-3} \bra{\psi_{0,0}} D^{(n)}_{da,db} \ket{\psi_{0,0}} D^{(n)}_{-da,-db}, \label{utv1}\\
	&\Pi_2 = \frac{(d-1)}{2d}\sum_{a,b=0}^{d-1} \bra{\psi_{0,0}} D^{(n)}_{(d-2)a,(d-2)b} \ket{\psi_{0,0}} D_{-(d-2)a,-(d-2)b}^{(n)}. \label{utv2}
\end{align}
See Appendix \ref{expansion}. 
The displacement operators that occur in these expansions are block-diagonal and, consequently, so are $\Pi_1$ and $\Pi_2$.
We can therefore rewrite Eqs.\ \eqref{first projection} and \eqref{second projection} as
\begin{align}
	&\Pi_1 = \frac{d-1}{2d} \sum_{j=1}^4 \sum_{a,b=0}^{d-1} 
			 D^{(n)}_{(d-2)a,(d-2)b} \Lambda_j\ketbra{\psi_{0,0}}{\psi_{0,0}}\Lambda_j D^{(n)}_{(2-d)a,(2-d)b}, \\
	&\Pi_2 = \frac{d-1}{2(d-2)} \sum_{j=1}^4 \sum_{a,b=0}^{d-3}
			 D^{(n)}_{da,db} \Lambda_j\ketbra{\psi_{0,0}}{\psi_{0,0}}\Lambda_j D^{(n)}_{-da,-db},
\end{align}
where $\Lambda_j$ is the orthogonal projection onto the $j$th subspace.
The operator $\Lambda_j\ketbra{\psi_{0,0}}{\psi_{0,0}}\Lambda_j$ can be regarded as an operator on the 
$j$th subspace, and, by \eqref{block1} and \eqref{block2},
the $j$th block of $\Pi_1$ and $\Pi_2$ can be written as
\begin{align}
	&\Pi_1^j = \frac{d-1}{2d} \sum_{a,b=0}^{d-1} 
			 (D^{(n_1)}_{-a,b}\otimes\1_{n_2}) \Lambda_j\ketbra{\psi_{0,0}}{\psi_{0,0}}\Lambda_j (D^{(n_1)}_{a,-b}\otimes\1_{n_2}), \\
	&\Pi_2^j = \frac{d-1}{2(d-2)}  \sum_{a,b=0}^{d-3}
			 (\1_{n_1}\otimes D^{(n_2)}_{a,b}) \Lambda_j\ketbra{\psi_{0,0}}{\psi_{0,0}}\Lambda_j (\1_{n_1}\otimes D^{(n_2)}_{-a,-b}).
\end{align}
The translation properties \eqref{disp trans prop} allow us to lower the upper limits of the sums:
\begin{align}
	&\Pi_1^j = \frac{(d-1)}{n_1} \sum_{a,b=0}^{n_1-1} 
			 (D^{(n_1)}_{-a,b}\otimes\1_{n_2}) \Lambda_j\ketbra{\psi_{0,0}}{\psi_{0,0}}\Lambda_j (D^{(n_1)}_{a,-b}\otimes\1_{n_2}), \\
	&\Pi_2^j = \frac{(d-1)}{n_2}  \sum_{a,b=0}^{n_2-1}
			 (\1_{n_1}\otimes D^{(n_2)}_{a,b}) \Lambda_j\ketbra{\psi_{0,0}}{\psi_{0,0}}\Lambda_j (\1_{n_1}\otimes D^{(n_2)}_{-a,-b}).
\end{align}
Finally, in Appendix \ref{Part Tr}, c.f.\ Eqs.\ \eqref{pt1} and \eqref{pt2}, we prove that these sums reduce to
\begin{align}
	\Pi_1^j &= (d-1) \1_{n_1} \otimes \tr_{n_1}(\Lambda_j\ketbra{\psi_{0,0}}{\psi_{0,0}}\Lambda_j), \label{partr1}\\
	\Pi_2^j &= (d-1) \tr_{n_2}(\Lambda_j\ketbra{\psi_{0,0}}{\psi_{0,0}}\Lambda_j)\otimes \1_{n_2}. \label{partr2}
\end{align}
The traces in \eqref{partr1} and \eqref{partr2} are the partial traces with respect to the 
splitting of the $j$th subspace as a tensor product.
We use the language of multipartite systems and refer to 
$(d-1) \tr_{n_1}(\Lambda_j\ketbra{\psi_{0,0}}{\psi_{0,0}}\Lambda_j)$ as the right marginal operator 
of $\Pi_1^j$ and to $(d-1) \tr_{n_2}(\Lambda_j\ketbra{\psi_{0,0}}{\psi_{0,0}}\Lambda_j)$ as the 
left marginal operator of $\Pi_2^j$.
In the next section we will prove that if the SIC is aligned, the right marginal operator of $\Pi_1^j$ 
is a projection operator, and we will calculate its rank. 
Then, since the two partial traces have the same spectrum (up to $0$s), 
the left marginal operator of $\Pi_2^j$ is also a projection operator, and it has the same rank.  

\subsection{Displaced parity operators}\label{Displaced parity operators}
Displaced parity operators will play an important role in our further 
analysis of the blocks of $\Pi_1$ and $\Pi_2$.
In this section we introduce these operators and describe some of their properties.

A parity operator is a Clifford unitary $P$ for which 
\begin{equation}\label{conjugation}
	PD_{a,b}^{(n)}P=D_{-a,-b}^{(n)}
\end{equation}
holds for all pairs of integers $a$ and $b$. Here, we have borrowed the terminology from crystallography; 
for an odd $n$, if you label the points in an $n$-periodic $2$-dimensional lattice by the indices of the displacement operators,
the action of $P$ corresponds to a reflection in the origin. For an even $n$, the analogy breaks down
due to the non-periodicity of the displacement operators, see Eq.\ \eqref{disp trans prop}. In Appendix \ref{Uniqueness of parity operators} we show that, irrespective 
of $n$ being odd or even,
there is (up to a sign) only one Clifford unitary which satisfies \eqref{conjugation}, namely
\begin{equation}\label{that guy}
	P^{(n)}=\sum_{u=0}^{n-1}\ketbra{-u}{u}.
\end{equation}
This may not be so surprising, considering the analogy with crystallography, 
but the proof is a good illustration of the difference in complexity between even and odd dimensions. 
The essential uniqueness justifies calling $P^{(n)}$ \emph{the parity operator}. 
In the definition \eqref{that guy}, $\{\ket{u}:u\in\ints_n\}$ is an orthonormal basis relative to which 
$X_n$ and $Z_n$ are represented as in Eq.\ \eqref{standard}.

The definition \eqref{conjugation} seems to 
depend on the representation of the Weyl-Heisenberg group. However, 
as was pointed out in Section \ref{clifford}, the Clifford groups associated with different representations
are canonically unitary equivalent, and the canonical isomorphism between the Clifford groups connects the two parity operators.
Therefore, the parity operator can be defined by \eqref{that guy} in any representation, although the basis on the righ-hand side is representation-dependent.

The parity operator is an involution. 
Recall that an involution is an operator which squares to the identity operator.
Involutions are diagonalizable and each eigenvalue equals either $+1$ or $-1$.
The multiplicities are determined by the trace of the involution; if $I$ is an involution on an $n$-dimensional 
Hilbert space, the multiplicity of the eigenvalue $+1$ is $(n+\tr I)/2$ and the multiplicity of $-1$ is 
$(n-\tr I)/2$. We write, for short, 
\begin{equation}\label{Iet}
	\spec I = \bigg(\frac{n+\tr I}{2},\frac{n-\tr I}{2}\bigg).
\end{equation}
The trace of the parity operator is $1$ if $n$ is odd and $2$ if $n$ is even.
Consequently,
\begin{equation}
	\spec P^{(n)} = 
	\begin{cases}
		\Big(\frac{n+1}{2},\frac{n-1}{2}\Big) & \text{if $n$ is odd}, \\
		\Big(\frac{n+2}{2},\frac{n-2}{2}\Big) & \text{if $n$ is even}.
	\end{cases}
\end{equation}
By displacing $P^{(n)}$ we can generate new involutions in the Clifford group:
\begin{equation}
	P^{(n)}_{a,b}=D_{a,b}^{(n)}P^{(n)}.
\end{equation}
If $n$ is odd, 
the displaced parity operators are unitarily equivalent to, and hence isospectral to, $P^{(n)}$. For in the odd case, $2k=a$ and $2l=b$ can 
always be solved in arithmetic modulo $n$, and by Eqs.\ \eqref{merging rule} and \eqref{conjugation}, $P^{(n)}_{a,b}=D_{k,l}^{(n)}P^{(n)}D_{-k,-l}^{(n)}$.
In the analogy with crystallography, $P^{(n)}_{a,b}$ corresponds to a reflection in the point $(k,l)$.
If $n$ is even, however, the situation is more complicated. In the even case, it is not only the identity operator that is preserved by the action of 
$P$, and the displaced parity operators divide into two unitary conjugacy classes. 
Irrespective of the parity of $n$ we have that
\begin{equation}
	\tr P^{(n)}_{a,b} 
	= \sum_{u=0}^{n-1}\tau^{ab}_n\bra{-u}X^aZ^b\ket{u}
	= \sum_{u=0}^{n-1}\tau^{ab+2bu}_n\delta_{2u+a,0}^{(n)},
\end{equation}
where $\delta^{(n)}_{\cdot,\cdot}$ is the Kronecker delta in arithmetic modulo $n$. Evaluation of the right-hand side for all possible values of $n$, $a$, and $b$ yields
\begin{equation}\label{notcalled}
	\tr P^{(n)}_{a,b} =
	\begin{cases}
		1 & \text{if $n$ is odd}, \\
		1-(-1)^{(a+1)(b+1)} & \text{if $n$ is even.}
	\end{cases}
\end{equation}
We see that, in the even case, the trace of a displaced parity operator can be $0$ or $2$.
If $\tr P^{(n)}_{a,b} = 2$, then
$a$ and $b$ have to be even, say $a=2k$ and $b=2l$, and 
$P^{(n)}_{a,b}=D_{k,l}^{(n)}P^{(n)}D_{-k,-l}^{(n)}$.
However, if $\tr P^{(n)}_{a,b} = 0$, then $P^{(n)}_{a,b}$
is not unitarily equivalent to $P^{(n)}$.
 
An immediate consequence of Eqs.\ \eqref{Iet} and \eqref{notcalled} is that
\begin{equation}\label{Seee}
	\spec P^{(n)}_{a,b} = 
	\begin{cases}
		\left(\frac{n+1}{2},\frac{n-1}{2}\right) & \text{if $n$ is odd}, \\
		\left(\frac{n+1-(-1)^{(a+1)(b+1)}}{2},\frac{n-1+(-1)^{(a+1)(b+1)}}{2}\right) & \text{if $n$ is even}.
	\end{cases}
\end{equation} 
Since the expansion of the parity operator in the displacement operator basis is
\begin{equation}
	P^{(n)} 
	= \frac{1}{n}\sum_{a,b=0}^{n-1} \tr\big(P_{a,b}^{(n)}\big) D_{-a,-b}^{(n)},
\end{equation}
we also conclude from \eqref{notcalled} that 
\begin{equation}\label{expansion of parity operator}
	P^{(n)} = 
	\begin{cases}
		\frac{1}{n}\sum_{a,b=0}^{n-1}D_{a,b}^{(n)} & \text{if $n$ is odd,} \\
		\frac{1}{n}\sum_{a,b=0}^{n-1}(1-(-1)^{(a+1)(b+1)})D_{a,b}^{(n)} & \text{if $n$ is even.}
	\end{cases}
\end{equation}
Equation \eqref{expansion of parity operator} is the key observation used in the next section.

\subsection{Proof that $\Pi_1$ is a projection operator}
So far, we have not used the assumption that the SIC is aligned.
In this section we will do so and calculate the blocks of $\Pi_1$.
More precisely, we will show that
\begin{align}
	&\Pi_1^1 = \frac{1}{2}\1_{n_1} \otimes ( \1_{n_2}\mp P^{(n_2)}_{0,0} ), \label{slut1}\\
	&\Pi_1^2 = \frac{1}{2}\1_{n_1} \otimes ( \1_{n_2}\pm P^{(n_2)}_{0,1} ), \label{slut2}\\
	&\Pi_1^3 = \frac{1}{2}\1_{n_1} \otimes ( \1_{n_2}\pm P^{(n_2)}_{-1,0} ), \label{slut3}\\
	&\Pi_1^4 = \frac{1}{2}\1_{n_1} \otimes ( \1_{n_2}\pm P^{(n_2)}_{-1,1} ). \label{slut4}
\end{align}
The upper signs are to be used if $n_2$ is odd and the lower signs are to be used if $n_2$ is even. 
Before that, however, let us consider some consequences of these identities.

\subsubsection{Consequences of Equations \eqref{slut1}-\eqref{slut4}}
Let us prove that the frames \eqref{frame1} and \eqref{frame2} are tight, given
that the blocks of $\Pi_1$ satisfy \eqref{slut1}-\eqref{slut4}.
Since the displaced parity operators are involutions, the blocks of $\Pi_1$, and hence $\Pi_1$ itself, are projection operators.
We calculate their ranks.

If $n_2$ is odd, then, by \eqref{Seee},
$\Pi_1^1$ has rank $n_1(n_2-1)/2$ while $\Pi_1^2$, $\Pi_1^3$, and $\Pi_1^4$ each have rank $n_1(n_2+1)/2$.
If $n_2$ is even, $\Pi_1^1$ has rank $n_1(n_2+2)/2$
while $\Pi_1^2$, $\Pi_1^3$, and $\Pi_1^4$ each have rank $n_1n_2/2$.
In either case,
\begin{equation}
	\Rk \Pi_1 = \frac{n_1(n_2-1)}{2} + \frac{3n_1(n_2+1)}{2} = \frac{n_1(n_2+2)}{2} + \frac{3n_1n_2}{2} = \frac{d(d-1)}{2}.
\end{equation}

Next we consider the operator $\Pi_2$.
Since the blocks of $\Pi_1$ are projection operators, so are the blocks of $\Pi_2$, as well as $\Pi_2$ itself;
for Eqs.\ \eqref{partr1} and \eqref{partr2} say that the left marginal 
operator of $\Pi_2^j$ has the same spectrum as the 
right marginal operator of $\Pi_1^j$. We conclude that if $n_2$ is odd, $\Pi_2^1$ has rank $n_2(n_2-1)/2$ while $\Pi_2^2$, $\Pi_2^3$, and $\Pi_2^4$ each have rank $n_2(n_2+1)/2$, and if $n_2$ is even, $\Pi_2^1$ has rank $n_2(n_2+2)/2$ while $\Pi_2^2$, $\Pi_2^3$, and $\Pi_2^4$ each have rank $n_2^2/2$. In either case, 
\begin{equation}
	\Rk \Pi_2 = \frac{n_2(n_2-1)}{2} + \frac{3n_2(n_2+1)}{2} = \frac{n_2(n_2+2)}{2} + \frac{3n_2^2}{2} = \frac{(d-1)(d-2)}{2}.
\end{equation}

We have shown that, under the assumption that Eqs.\ \eqref{slut1}-\eqref{slut4} hold,
an aligned SIC in dimension $d(d-2)$ contains a tight $d^2$-frame of rank $d(d-1)/2$
and a tight $(d-2)^2$-frame of rank $(d-1)(d-2)/2$. 
By displacing these frames we will generate the whole SIC.
In other words, the SIC 
consists of $(d-2)^2$ tight $d^2$-frames, and, alternatively,
of  $d^2$ tight $(d-2)^2$-frames. In the following section we expand on the
proof of the structure of $\Pi_1$.
Afterwards we discuss implications on the symmetry of aligned SICs.

\subsubsection{Derivations of Equations \eqref{slut1}-\eqref{slut4}}

Using definition \eqref{overlap phases}, the expansion \eqref{utv1} can be rearranged as
\begin{equation}\label{pipi1}
		\Pi_1 = \frac{1}{2}\1_{n_2}+\frac{1}{4n_2} \sum_{a,b=0}^{d-3} e^{i\theta_{da,db}^{(n)}} D^{(n)}_{-da,-db}.
\end{equation}
Then, by \eqref{block1}, 
the blocks of $\Pi_1$ are given by
\begin{align}
	\Pi_1^1 &= \frac{1}{2}\1_{n_1} \otimes \Big(
		\1_{n_2} + \frac{1}{2n_2}\sum_{a,b=0}^{d-3}e^{i\theta_{da,db}^{(n)}} D_{a,b}^{(n_2)}\Big), \label{oddeq1}\\
	\Pi_1^2 &= \frac{1}{2}\1_{n_1} \otimes \Big(
		\1_{n_2} + \frac{1}{2n_2}\sum_{a,b=0}^{d-3}e^{i\theta_{da,db}^{(n)}} \omega_{2n_2}^a  D_{a,b}^{(n_2)}\Big), \label{oddeq2}\\
	\Pi_1^3 &= \frac{1}{2}\1_{n_1} \otimes \Big(
		\1_{n_2} + \frac{1}{2n_2}\sum_{a,b=0}^{d-3}e^{i\theta_{da,db}^{(n)}} 
		\omega_{2n_2}^b  D_{a,b}^{(n_2)} \Big), \label{oddeq3}\\
	\Pi_1^4 &= \frac{1}{2}\1_{n_1} \otimes \Big(
		\1_{n_2} + \frac{1}{2n_2}\sum_{a,b=0}^{d-3}e^{i\theta_{da,db}^{(n)}} 
		\omega_{2n_2}^{a+b}  D_{a,b}^{(n_2)} \Big). \label{oddeq4}
\end{align}
We will now prove that, under the alignment assumption \eqref{alignment condition1}, these expressions equal those in Eqs.\ \eqref{slut1}-\eqref{slut4}.

According to the alignment assumption, the overlap phases 
for the displacement operators appearing in the expansion \eqref{pipi1} of $\Pi_1$ 
are 
\begin{equation}
	e^{i\theta_{da,db}^{(n)}} = -(-1)^{(a+1)(b+1)}.
\end{equation}
(Notice that this formula holds if $a=0\md (d-2)$ and $b = 0\md (d-2)$ as well).
The overlap phases satisfy the translation properties  
\begin{equation}
	e^{i\theta_{d(a+m),db}^{(n)}} =
	\begin{cases}
		e^{i\theta_{da,db}^{(n)}} & \text{ if $m$ is even}, \\
		e^{i\theta_{d(a+1),db}^{(n)}} & \text{ if $m$ is odd},
	\end{cases}
\end{equation}
and 
\begin{equation}
	e^{i\theta_{da,d(b+m)}^{(n)}} =
	\begin{cases}
		e^{i\theta_{da,db}^{(n)}} & \text{ if $m$ is even}, \\
		e^{i\theta_{da,d(b+1)}^{(n)}} & \text{ if $m$ is odd}.
	\end{cases}
\end{equation}
Using these, the translation properties \eqref{disp trans prop},
and the identity $\omega_{2n_2}^{m+n_2} = -\omega_{2n_2}^{m}$,
one can reduce the upper limits in the sums in Eqs.\ \eqref{oddeq1}-\eqref{oddeq4} to $n_2-1$.
More precisely, one can show that if $n_2$ is odd, then
\begin{align}
	&\sum_{a,b=0}^{d-3} e^{i\theta_{da,db}^{(n)}} D_{a,b}^{(n_2)} 
		= -2\sum_{a,b=0}^{n_2-1} D_{a,b}^{(n_2)}, \label{odd1}\\
	&\sum_{a,b=0}^{d-3} e^{i\theta_{da,db}^{(n)}} \omega_{2n_2}^{a} D_{a,b}^{(n_2)} 
		= 2\sum_{a,b=0}^{n_2-1} \tau_{n_2}^{a} D_{a,b}^{(n_2)}, \label{odd2}\\
	&\sum_{a,b=0}^{d-3} e^{i\theta_{da,db}^{(n)}} \omega_{2n_2}^{b} D_{a,b}^{(n_2)} 
		= 2\sum_{a,b=0}^{n_2-1} \tau_{n_2}^{b} D_{a,b}^{(n_2)}, \label{odd3}\\
	&\sum_{a,b=0}^{d-3} e^{i\theta_{da,db}^{(n)}} \omega_{2n_2}^{a+b} D_{a,b}^{(n_2)} 
		= 2\sum_{a,b=0}^{n_2-1} \tau_{2n_2}^{a+b} D_{a,b}^{(n_2)}, \label{odd4}
\end{align}
and if $n_2$ is even,
\begin{align}
	&\sum_{a,b=0}^{d-3} e^{i\theta_{da,db}^{(n)}} D_{a,b}^{(n_2)} 
		= \sum_{a,b=0}^{n_2-1}(1-(-1)^{(a+1)(b+1)}) D_{a,b}^{(n_2)}, \label{even1}\\
	&\sum_{a,b=0}^{d-3} e^{i\theta_{da,db}^{(n)}} \omega_{2n_2}^{a} D_{a,b}^{(n_2)} 
		= \sum_{a,b=0}^{n_2-1}((-1)^{a}+1)((-1)^{b}-1)\tau_{n_2}^{a} D_{a,b}^{(n_2)}, \label{even2}\\
	&\sum_{a,b=0}^{d-3} e^{i\theta_{da,db}^{(n)}} \omega_{2n_2}^{b} D_{a,b}^{(n_2)} 
		= \sum_{a,b=0}^{n_2-1}((-1)^{a}-1)((-1)^{b}+1)\tau_{n_2}^{b} D_{a,b}^{(n_2)}, \label{even3}\\
	&\sum_{a,b=0}^{d-3} e^{i\theta_{da,db}^{(n)}} \omega_{2n_2}^{a+b} D_{a,b}^{(n_2)} 
		= -\sum_{a,b=0}^{n_2-1}((-1)^{a}-1)((-1)^{b}-1)\tau_{n_2}^{a+b} D_{a,b}^{(n_2)}. \label{even4}
\end{align}
Equation \eqref{slut1} follows immediately from \eqref{oddeq1} 
and a comparison between Eqs.\ \eqref{expansion of parity operator} and \eqref{odd1} in the odd case,
and between Eqs.\ \eqref{expansion of parity operator} and \eqref{even1} in the even case.

Next, we consider the Eqs.\ \eqref{odd2} and \eqref{even2}.
If $n_2$ is odd, then
\begin{equation}\label{evk1}
	\sum_{a,b=0}^{n_2-1} \tau_{n_2}^{a} D_{a,b}^{(n_2)}
	= \sum_{a,b=0}^{n_2-1} D_{a,b+1}^{(n_2)} Z_{n_2}^{-1}
	= \sum_{a,b=0}^{n_2-1} D_{a,b}^{(n_2)} Z_{n_2}^{-1}	
	= n_2P^{(n_2)}Z_{n_2}^{-1}
	= n_2P^{(n_2)}_{0,1}.
\end{equation}
The second identity follows from the translation property \eqref{disp trans prop}, the 
third from Eq.\ \eqref{expansion of parity operator}, and the fourth from \eqref{conjugation}.
If $n_2$ is even, then
\begin{equation}\label{sextifem}
\begin{split}
	\sum_{a,b=0}^{n_2-1} ((-1)^{a}+1)((-1)^{b}-1)  \tau_{n_2}^{a} D_{a,b}^{(n_2)} 
	&= \sum_{a,b=0}^{n_2-1} ((-1)^{a}+1)((-1)^{b}-1) D_{a,b+1}^{(n_2)} Z_{n_2}^{-1} \\
	&= -2\sum_{a,b=0}^{n_2-1} (1-(-1)^{(a+1)(b+1)})  D_{a,b}^{(n_2)} Z_{n_2}^{-1}.
\end{split}
\end{equation} 
Again, in the second identity we used \eqref{disp trans prop}, and we rewrote the factors in front of the displacement operators. Using Eqs.\ \eqref{expansion of parity operator} and \eqref{conjugation}, we identify the right-hand side of \eqref{sextifem} as $-2n_2P^{(n_2)}_{0,1}$. This finishes the proof of Eq.\ \eqref{slut2}.
The proofs of Eqs.\ \eqref{slut3} and \eqref{slut4} are similar to the proof of \eqref{slut2} and, hence, we omit them. 

\section{Symmetry}\label{sec4}
By a symmetry of a SIC we mean any unitary which permutes the SIC vectors.
In this section we show that any aligned WH-SIC in dimension $n=d(d-2)$, where $d$ is even,
has a symplectic symmetry of order $2$ which leaves unchanged a SIC fiducial satisfying the alignment condition \eqref{alignment condition1}. The corresponding result 
for $d$ odd was proven in \cite{ApBeDuFl2017}.
 
In Section \ref{block-diagonal splitting} we have shown that the Hilbert space can be decomposed into four 
subspaces, each admitting a tensor product splitting relative to which the blocks of $\Pi_1$ 
acquire the form in Eq.\ \eqref{partr1}. It follows from Eq.\ \eqref{partr1} and 
Eqs.\ \eqref{slut1}-\eqref{slut4} that 
\begin{equation}\label{denna}
	\tr_{n_1}(\Lambda_j\ketbra{\psi_{0,0}}{\psi_{0,0}}\Lambda_j) = \frac{1}{2(d-1)}( \1_{n_2} + P_j )
\end{equation}
where 
\begin{align}
	&P_1 = \mp P^{(n_2)}_{0,0},
	&&P_2 = \pm P^{(n_2)}_{0,1},
	&&P_3 = \pm P^{(n_2)}_{-1,0},
	&&P_4 = \pm P^{(n_2)}_{-1,1}.
\end{align}
Recall that the upper signs are to be used if $n_2$ is odd and the lower signs are to be used if $n_2$ is even. We fix an orthonormal 
basis $\{\ket{f_u;j}:u\in\ints_{n_2}\}$ in the second factor of the $j$th subspace which diagonalizes $P_j$ in such a way that its eigenvalues are arranged in descending order:
\begin{equation}
	P_j = \sum_{u=0}^{m_j-1} \ketbra{f_u;j}{f_u;j} - \sum_{u=0}^{n_2-m_j}\ketbra{f_u;j}{f_u;j}.
\end{equation}
The upper limits are determined by Eq.\ \eqref{Seee}. That is, 
\begin{equation}
m_1=\begin{cases}
	(n_2-1)/2 & \text{if $n_2$ is odd,}\\
	(n_2-1)/2 & \text{if $n_2$ is even,}
	\end{cases}
\end{equation}
and 
\begin{equation}
m_2=m_3=m_4=\begin{cases}
	(n_2+1)/2 & \text{if $n_2$ is odd,}\\
	n_2/2 & \text{if $n_2$ is even.}
	\end{cases}
\end{equation}
The diagonalizing bases for the parity operators can be completed
to Schmidt-bases for the projections of the SIC fiducial \cite{EkKn1995}.
According to Eq.\ \eqref{denna} there thus exist mutually orthogonal unit vectors 
$\ket{e_u;j}$ in the first factor in the $j$th subspace such that 
 \begin{equation}
	\Lambda_j\ket{\psi_{0,0}}= \frac{1}{\sqrt{d-1}} 
	 \sum_{u=0}^{m_j} \ket{e_u;j}\otimes\ket{f_u;j}.
\end{equation}
Define a unitary $U_b$ by
\begin{equation}
	U_b
 	= \begin{pmatrix}
		\1_{n_1} \otimes P^{(n_2)}_{0,0} & & & \\
		 & -\1_{n_1} \otimes P^{(n_2)}_{0,1} & & \\
		 & & -\1_{n_1} \otimes P^{(n_2)}_{-1,0} & \\
		 & & & -\1_{n_1} \otimes P^{(n_2)}_{-1,1} 
	  \end{pmatrix}. \label{symmetry}
\end{equation}
The unitary clearly leaves the SIC fiducial unchanged and is of second order, since the parity operators are of second order. If, in addition, $U_b$ permutes the other SIC vectors, it is a symmetry.
This is the case if $U_b$ belongs to the Clifford group. 
We next prove that $U_b$ is, in fact, the symplectic unitary corresponding to 
\begin{equation}
	F_b = \begin{pmatrix}
			1-d & n \\ n & 1-d + n
		\end{pmatrix} 
	  = \begin{pmatrix}
			-n & 1-d \\ d-1-n & n
		\end{pmatrix} 
		\begin{pmatrix}
			0 & -1 \\ 1 & 0
		\end{pmatrix},
\end{equation}
the product in the right-hand side being a prime decomposition of $F_b$ in $\SL(2,\ints_{\bar n})$.
The choice of symplectic matrix is 
inspired by a conjecture of Scott and Grassl \cite{ScGr2010, Sc2017}.

The inverse of $1-d$ modulo $\bar n$ is $(1-d)(n+1)$.
Applying \eqref{tut} and \eqref{the rep} yields 
\begin{equation}
	V_{F_b}=\sum_{u=0}^{n-1} (-1)^u\ketbra{u}{dn_2-(d-1)u}.
\end{equation}
The expansion of $V_{F_b}$ in the displacement operator basis then reads
\begin{equation}
\begin{split}
	V_{F_b}
	&= \frac{1}{n} \sum_{a,b=0}^{n-1} \sum_{u=0}^{n-1} (-1)^u \bra{dn_2-(d-1)u} D_{a,b}^{(n)} \ket{u} D_{-a,-b}^{(n)} \\
	&= \frac{1}{n} \sum_{a,b=0}^{n-1} \sum_{u=0}^{n-1} (-1)^u \tau_n^{ab} \omega_n^{bu} \braket{dn_2-(d-1)u}{u+a}  D_{-a,-b}^{(n)}\\
	&= \frac{1}{n} \sum_{a,b=0}^{n-1} \sum_{u=0}^{n-1} (-1)^u \tau_n^{ab} \omega_n^{bu} \delta^{(n)}_{a,dn_2-du} D_{-a,-b}^{(n)}.
\end{split}
\end{equation}
The Kronecker delta is non-zero only if $a$ is divisible by $d$ and $u=n_2-a/d\md 2n_2$.
Hence, we can rewrite the expansion of $V_{F_b}$ as 
\begin{equation}
\begin{split}
	V_{F_b}
	&= \frac{1}{n} \sum_{a=0}^{d-3} \sum_{b=0}^{n-1} \sum_{k=0}^{d-1} (-1)^u \tau_n^{dab} \omega_n^{bu} \delta^{(n)}_{a,n_2-u} D_{-da,-b}^{(n)}  \\
	&= \frac{1}{n} \sum_{a=0}^{n_2} \sum_{b=0}^{n-1} \sum_{k=0}^{d-1} 
	(-1)^{n_2-a} \tau_n^{dab} \omega_n^{b(n_2-a)} \omega_n^{2bn_2k} D_{-da,-b}^{(n)}	\\
	&\qquad + \frac{1}{n} \sum_{a=n_1}^{d-3} \sum_{b=0}^{n-1} \sum_{k=1}^{d} 
	(-1)^{n_2-a} \tau_n^{dab} \omega_n^{b(n_2-a)} \omega_n^{2bn_2k} D_{-da,-b}^{(n)} \\
	&= \frac{1}{d-2} \sum_{a=0}^{n_2} \sum_{b=0}^{n-1} 
	(-1)^{n_2-a} \tau_n^{dab} \omega_n^{b(n_2-a)} \delta_{b,0}^{(d)} D_{-da,-b}^{(n)}	\\
	&\qquad + \frac{1}{d-2} \sum_{a=n_1}^{d-3} \sum_{b=0}^{n-1} 
	(-1)^{n_2-a} \tau_n^{dab} \omega_n^{b(n_2-a)} \omega_d^{b} \delta_{b,0}^{(d)} D_{-da,-b}^{(n)} \\
	&= \frac{1}{d-2} \sum_{a=0}^{d-3} \sum_{b=0}^{n-1} 
	(-1)^{n_2-a} \tau_n^{dab} \omega_n^{b(n_2-a)} \delta_{b,0}^{(d)} D_{-da,-b}^{(n)}.
\end{split}
\end{equation}
Only those terms in which $b$ is divisible by $d$ are thus non-zero and, hence, 
\begin{equation}
\begin{split}
	V_{F_b}
	&= \frac{1}{d-2} \sum_{a,b=0}^{d-3}
	(-1)^{n_2-a} \tau_n^{d^2ab} \omega_n^{db(n_2-a)} D_{-da,-db}^{(n)}\\
	&=\frac{1}{d-2} \sum_{a,b=0}^{d-3}
	(-1)^{n_2+a+b+ab} D_{-da,-db}^{(n)}.
\end{split}
\end{equation}
According to Eq.\ \eqref{block1}, $V_{F_b}$ is block-diagonal and the blocks split:
\begin{equation}
	V_{F_b}
	= \frac{1}{d-2} \sum_{a,b=0}^{d-3} (-1)^{n_2+a+b+ab} 
		\begin{pmatrix} 
		\1_{n_1} \otimes D_{-a,-b}^{(n_2)} & & & \\
		 & \hspace{-38pt}\1_{n_1} \otimes \omega_{2n_2}^{-a}  D_{-a,-b}^{(n_2)} & & \\
		 & & \hspace{-38pt}\1_{n_1} \otimes \omega_{2n_2}^{-b}  D_{-a,-b}^{(n_2)} & \\
		 & & & \hspace{-38pt}\1_{n_1} \otimes \omega_{2n_2}^{-(a+b)}  D_{-a,-b}^{(n_2)} 
	  	\end{pmatrix}.
\end{equation}
Direct calculations using the translation properties \eqref{disp trans prop} yield that 
if $n_2$ is odd,
\begin{align}
	& \sum_{a,b=0}^{d-3} (-1)^{n_2+a+b+ab} D_{-a,-b}^{(n_2)} 
		= 2 \sum_{a,b=0}^{n_2-1} D_{-a,-b}^{(n_2)}, \label{attifem}\\
	&\sum_{a,b=0}^{d-3} (-1)^{n_2+a+b+ab} \omega_{2n_2}^{-a}  D_{-a,-b}^{(n_2)} 
		= -2 \sum_{a,b=0}^{n_2-1} \tau_{n_2}^{-a} D_{-a,-b}^{(n_2)}, \\
	&\sum_{a,b=0}^{d-3} (-1)^{n_2+a+b+ab} \omega_{2n_2}^{-b} D_{-a,-b}^{(n_2)} 
		= -2 \sum_{a,b=0}^{n_2-1} \tau_{n_2}^{-b} D_{-a,-b}^{(n_2)}, \\
	&\sum_{a,b=0}^{d-3} (-1)^{n_2+a+b+ab} \omega_{2n_2}^{-(a+b)} D_{-a,-b}^{(n_2)} 
		= -2 \sum_{a,b=0}^{n_2-1} \tau_{n_2}^{-(a+b)} D_{-a,-b}^{(n_2)},
\end{align}
and if $n_2$ is even,
\begin{align}
	&\sum_{a,b=0}^{d-3} (-1)^{n_2+a+b+ab} D_{-a,-b}^{(n_2)} 
		= 2 \sum_{a,b=0}^{n_2-1} (1-(-1)^{(a+1)(b+1)})D_{-a,-b}^{(n_2)},  \label{attinio}\\
	&\sum_{a,b=0}^{d-3} (-1)^{n_2+a+b+ab} \omega_{2n_2}^{-a} D_{-a,-b}^{(n_2)} 
		= -\sum_{a,b=0}^{n_2-1} (1+(-1)^a)(1-(-1)^b)\omega_{2n_2}^{-a} D_{-a,-b}^{(n_2)}, \\
	&\sum_{a,b=0}^{d-3} (-1)^{n_2+a+b+ab} \omega_{2n_2}^{-b} D_{-a,-b}^{(n_2)} 
		= - \sum_{a,b=0}^{n_2-1} (1-(-1)^a)(1+(-1)^b) \omega_{2n_2}^{-b} D_{-a,-b}^{(n_2)}, \\
	&\sum_{a,b=0}^{d-3} (-1)^{n_2+a+b+ab} \omega_{2n_2}^{-(a+b)}  D_{-a,-b}^{(n_2)} 
		= - \sum_{a,b=0}^{n_2-1} (1-(-1)^a)(1-(-1)^b) \omega_{2n_2}^{-(a+b)} D_{-a,-b}^{(n_2)}.
\end{align}
The right-hand sides in \eqref{attifem} and \eqref{attinio} equal $2n_2 P_{0,0}^{(n_2)}$, c.f.\ Eq.\ \eqref{expansion of parity operator}, and, hence, the first block of $V_{F_b}$ is
\begin{equation}
	\frac{1}{d-2} \sum_{a,b=0}^{d-3} (-1)^{n_2+a+b+ab} \1_{n_1} \otimes D_{-a,-b}^{(n_2)} 
	= \1_{n_1} \otimes P^{(n_2)}_{0,0}.
\end{equation}
Furthermore, by a comparison with Eqs.\ \eqref{evk1} and \eqref{sextifem}, we see that, irrespective of the parity of $n_2$,
the second block of $V_{F_b}$ is
\begin{equation}
	\frac{1}{d-2} \sum_{a,b=0}^{d-3} (-1)^{n_2+a+b+ab}  \1_{n_1} \otimes \omega_{2n_2}^{-a} D_{-a,-b}^{(n_2)} 
	= - \1_{n_1} \otimes P^{(n_2)}_{0,1}.
\end{equation}
Similarly, one can show that the third and fourth blocks of $V_{F_b}$ are
\begin{align}
	& \frac{1}{d-2} \sum_{a,b=0}^{d-3} (-1)^{n_2+a+b+ab} \1_{n_1} \otimes \omega_{2n_2}^{-a} D_{-a,-b}^{(n_2)} 
	= - \1_{n_1} \otimes P^{(n_2)}_{-1,0}, \\
	&\frac{1}{d-2} \sum_{a,b=0}^{d-3} (-1)^{n_2+a+b+ab} \1_{n_1} \otimes \omega_{2n_2}^{-(a+b)} D_{-a,-b}^{(n_2)} 
	= - \1_{n_1} \otimes P^{(n_2)}_{-1,1},
\end{align}
respectively. This proves that $U_b=V_{F_b}$.

\section{Conclusion}
We have proven that the property of alignment of WH-SICs in even
dimensions of the form $d(d-2)$ implies that the SICs can be partitioned into sets of 
equiangular tight frames, in two different ways. Together with \cite{ApBeDuFl2017}, which proves the same
for  SICs in odd dimensions of the form $d(d-2)$, this 
concludes the proof of the implication for all aligned WH-SICs.

The proof in \cite{ApBeDuFl2017} employs a powerful tool for handling the Weyl-Heisenberg group in composite
dimensions, namely Chinese remaindering.
In the past, Chinese remaindering has only been successfully used for Hilbert spaces of composite dimensions 
where the factors are relatively prime. In this paper, we have used special properties of irreducible representations of the Weyl-Heisenberg group in dimensions 
divisible by $4$ to decompose the Hilbert space into four subspaces, and to apply Chinese remaindering in each of them. Thus we have extended the use of Chinese remaindering to 
composite dimensions where the factors are not relatively prime. 
A generalization of our procedure to all composite dimensions is not immediately available.
However, decomposing the Hilbert space into a direct sum presents itself as a natural 
tool for tackling composite dimensions with Chinese remaindering, and it will be interesting to see whether 
it can be employed in other cases. 

Finally, we have proved that an extra symmetry, conjectured for aligned SICs and proven in \cite{ApBeDuFl2017} for the odd-dimensional case,
is indeed always present in the aligned SICs.

\appendix

\section{An unorthodox representation of the Weyl-Heisenberg group}\label{unusual}
In this appendix, we prove that if the dimension of the Hilbert space is divisible by $4$, 
the space can be decomposed into $4$ subspaces in such a way that the displacement operators with even indices 
assume the block-diagonal form in Eq.\ \eqref{dispdiag}.

Let $\HH^n$ be an $n$-dimensional Hilbert space.
Assume that $n$ is divisible by $4$ and write $n=4m$.
Fix an orthonormal basis 
$\{\ket{u;j}:u\in\ints_m, j=1,\dots,4\}$
for $\HH^n$, which we assume to be lexicographically ordered, and write $\HH_j^m$ for the linear span of $\{\ket{u;j}:u\in\ints_m\}$.
Furthermore, define operators $\1_m^{ji}$, $X_m^{ji}$, and $Z_m^{ji}$ from $\HH^m_i$ onto $\HH^m_{j}$ by
\begin{align}
	&\1_m^{ji} =	\sum_{u=0}^{m-1}\ketbra{u;j}{u;i}, 				
	&&X_m^{ji}	 =	\sum_{u=0}^{m-1}\ketbra{u+1;j}{u;i}, 
	&&Z_m^{ji}	 =	\sum_{u=0}^{m-1}\omega_m^{u}\ketbra{u;j}{u;i},
\end{align}
and let $\Lambda_j$ be the orthogonal projection of $\HH^n$ onto $\HH^n_j$.

The operators $X_m^{jj}$ and $Z_m^{jj}$ define irreducible representations of $\WH(m)$ on $\HH_j^m$. 
Inspired by \cite{ApBeBrErGrLa2012}, we define an $m$-nomial unitary representation 
of $\WH(n)$ on $\HH^n$ by declaring that 
\begin{align}\label{declaration}
	&X_n	=	\begin{pmatrix}
			  	0 			& 0  		& X_m^{13} 	& 0  					\\
			 	0 			& 0 		& 0  		& \omega_{2m}X_m^{24}	\\
				\1_m^{31} 	& 0  		& 0  		& 0 					\\
			 	0 			& \1_m^{42} & 0  		& 0 
		 	\end{pmatrix},
	&&Z_n	=	\begin{pmatrix}
			  	0			& \1_m^{12} & 0  					& 0						\\
				Z_m^{21} 	& 0  		& 0  					& 0						\\
			  	0			& 0  		& 0  					& \omega_{4m}\1_m^{34}  \\
			  	0			& 0			& \omega_{4m}Z_m^{43} 	& 0  
		 	\end{pmatrix}.
\end{align}
In these matrix representations,
the operators on position $(j,i)$ correspond to $\Lambda_{j} X_n \Lambda_i$ and 
$\Lambda_{j} Z_n \Lambda_i$, respectively, regarded as operators from $\HH_i^m$ to $\HH_{j}^m$. 
Below we will show that the representation defined by Eq.\ \eqref{declaration} is irreducible.
But before we do that, let us emphasize an important feature of the representation and 
discuss one crucial implication which is key in this paper.

A straightforward calculation shows that the displacement operators on $\HH^n$
(i.e, those associated with the representation in \eqref{declaration})
with even indices are block-diagonal with respect to the decomposition 
of $\HH^n$ into the four mutually orthogonal subspaces $\HH_j^m$:
\begin{equation}\label{blockus}
	D^{(n)}_{2a,2b}	
	= (-1)^{ab}\begin{pmatrix}
		D^{(m;1)}_{a,b} & & & \\
		 & \omega_m^{a}D^{(m;2)}_{a,b} & & \\
		 & & \omega_m^{b}D^{(m;3)}_{a,b} & \\
		 & & & \omega_m^{a+b}D^{(m;4)}_{a,b}
		 \end{pmatrix}.
\end{equation}
The displacement operator $D_{a,b}^{(m;j)}$ in the right-hand side 
is the displacement operation associated with the representation of $\WH(m)$
on $\HH_j^m$ specified by $X_m^{jj}$ and $Z_m^{jj}$.
Then, by unitary equivalence, see Sec.\ \ref{invariance},
for \emph{any} irreducible representation of $\WH(n)$ on $\HH^n$
there exists a decomposition of $\HH^n$ into four mutually orthogonal 
$m$-dimensional subspaces, and irreducible representations of $\WH(m)$ on these subspaces,
such that the displacement operators with even indices of the $\WH(n)$ 
representation assume a block-diagonal form like in \eqref{blockus}.
 
We will now prove that the representation specified by Eq.\ \eqref{declaration} is irreducible.
We do this by proving that it is unitarily equivalent to the `standard' representation of $\WH(n)$, in which the unitary operators corresponding to $X$ and $Z$ are represented by generalized Pauli matrices, c.f.\ Eq.\
\eqref{standard}. To this end we introduce, for any integer $s\ge 2$, two $s\times s$ matrices 
\begin{align}\label{matriklar}
	&\XXX_s	=	\begin{pmatrix}
			0 		& 0 		& \!\cdots\! 	& 0 & 1		\\
			1 		& 0	& \!\cdots\! 	& 0 & 0 		\\
			0 		& 1 		& \!\cdots\! 	& 0 & 0		\\
			\vdots 	& \vdots 	&  				& \vdots & \vdots 	\\
		 	0 		& 0 		& \!\cdots\! 	& 1 & 0
				\end{pmatrix},
	&&\ZZZ_s	=	\begin{pmatrix}
			1 		& 0 		& \!\cdots\! 	& 0		\\
			0 		& \omega_s	& \!\cdots\! 	& 0 		\\
			\vdots 	& \vdots 	&  				& \vdots 	\\
		 	0 		& 0 		& \!\cdots\! 	& \omega_s^{s-1}
			\end{pmatrix},
\end{align}
where, as usual, $\omega_s=e^{2\pi i/s}$.
We also introduce two unitary $2s\times 2s$ matrices 
\begin{align}
	&	\VVV_{2s} = \begin{pmatrix} \VVV & \mathbf{0} \\ \mathbf{0} & \VVV \end{pmatrix},
	&&	\WWW_{2s} =	\begin{pmatrix} \FFF_s & 0 \\ 0 & \FFF_s \end{pmatrix}.
\end{align}
The bold zeroes denote columns of $(s-1)$ zeros, and $\VVV$ and $\FFF_s$ are the $s\times (2s-1)$ matrix and the $s\times s$ matrix, respectively, given by 
\begin{align}\label{matriklarna}
	&\VVV = \begin{pmatrix} 
			1 	   & 0 		& 0 	 & 0 	  & 0 	   & \cdots & 0 	 \\ 
			0 	   & 0 		& 1 	 & 0 	  & 0 	   & \cdots & 0 	 \\
			0 	   & 0 		& 0 	 & 0 	  & 1 	   & \cdots & 0 	 \\
			\vdots & \vdots & \vdots & \vdots & \vdots &  		& \vdots \\
			0 	   & 0 		& 0 	 & 0	  &	0 	   & \cdots & 1 	 	
			\end{pmatrix},
	&\FFF_s = \begin{pmatrix} 
			1 	   & 1 				& 1 	  	   		& \cdots & 1 	 				\\ 
			1 	   & \omega_s 		& \omega_s^2 	 	& \cdots & \omega_s^{s-1} 	 	\\
			1 	   & \omega_s^2 	& \omega_s^4 	 	& \cdots & \omega_s^{2(s-1)} 	\\
			\vdots & \vdots 		& \vdots 			&  		 & \vdots 				\\
			1 	   & \omega_s^{s-1} & \omega_s^{2(s-1)} & \cdots & \omega_s^{(s-1)^2} 	 	
			\end{pmatrix}.	
	\end{align}
The matrix $\VVV_{2s}$ satisfies
\begin{align}
	&\VVV_{2s} \begin{pmatrix} 0 & \XXX_s \\ \III_{s} & 0 \end{pmatrix} \VVV_{2s}^\dagger = \XXX_{2s},
	&&\VVV_{2s} \begin{pmatrix} \ZZZ_s & 0 \\  0 & \omega_{2s}\ZZZ_s \end{pmatrix} \VVV_{2s}^\dagger  = \ZZZ_{2s},
\end{align}
where $\III_{s}$ is the $s\times s$ identity matrix. Moreover, the matrix $\FFF_s$, which is the discrete $s\times s$ Fourier transform, satisfies
\begin{align}
& \FFF_s\XXX_s\FFF_s^\dagger=\ZZZ_s,
&&\FFF_s\ZZZ_s\FFF_s^\dagger=\XXX_s^\dagger.
\end{align}
To prove the second equality, first note that the square of the Fourier transform is the parity operator, see Equation \eqref{that guy}, and then use the property \eqref{conjugation}.
The matrix $\WWW_{2s}$ satisfies
\begin{align}
	&\WWW_{2s} \begin{pmatrix} \XXX_s & 0 \\ 0 & \omega_{2s}\XXX_s \end{pmatrix} \WWW_{2s}^\dagger 
		= \begin{pmatrix} \ZZZ_s & 0 \\ 0 & \omega_{2s}\ZZZ_s \end{pmatrix},
	&&\WWW_{2s} \begin{pmatrix} 0 & \III_{s} \\ \ZZZ_s  & 0\end{pmatrix} \WWW_{2s}^\dagger 
		= \begin{pmatrix}0 & \III_{s} \\ \XXX_s^\dagger  & 0\end{pmatrix}.
\end{align}
The unitary $\UUU_n$, defined as
\begin{equation}
	\UUU_n=\VVV_{4m} \begin{pmatrix} \FFF_{2m}^\dagger \VVV_{2m} \WWW_{2m} & 0
	\\ 0 & \FFF_{2m}^\dagger \VVV_{2m} \WWW_{2m} \end{pmatrix},
\end{equation}
is then such that

\begin{align}\label{Xasd}
	&\UUU_n 	\begin{pmatrix}
				 0 & 0  & \XXX_m & 0  \\
				 0 & 0  &  0 & \omega_{2m}\XXX_m \\
				\III_m & 0  &  0 &  0 \\
				 0 & \III_m &  0 & 0  
		  	\end{pmatrix} \UUU_n^\dagger = \XXX_{n},
	&&\UUU_n	\begin{pmatrix}
				  0 & \III_m & 0  &  0 \\
				\ZZZ_m & 0  &  0 &  0 \\
				  0 & 0  & 0  & \omega_{4m}\III_m \\
				  0 & 0  & \omega_{4m}\ZZZ_m & 0  
		 	\end{pmatrix} \UUU_n^\dagger = \ZZZ_{n}.
\end{align}
We let $U_n$ be the unitary operator on $\HH^n$ which is represented by the matrix
$\UUU_n$ relative to the chosen basis for $\HH^n$.
By Eqs.\ \eqref{declaration} and \eqref{Xasd}, $U_nX_nU_n^\dagger$ and $U_nZ_nU_n^\dagger$ 
are represented by generalized Pauli matrices.

\section{Chinese remaindering}\label{CRT}
In this appendix, we present an application of the classic Chinese Remainder Theorem to representations of the Weyl-Heisenberg group. D. Gross, who came up with the idea, called the application ``Chinese remaindering'' \cite{Gross}. Hence the title of the appendix.
The presentation is inspired by \cite{Apunpub}.

Let $n_1$ and $n_2$ be two positive and relatively prime integers
and set $m=n_1n_2$.
The Chinese Remainder Theorem states that the rings $\ints_{m}$ and 
$\ints_{n_1}\times\ints_{n_2}$ are isomorphic. An isomorphism is given by
\begin{equation}
	u\md m\to (u\md n_1,u\md n_2).
\end{equation}
For simplicity, we will write $u$ for $u\md m$
and, then, write $u_1$ for $u\md n_1$ and $u_2$ for $u\md n_2$.
We also define $\bar m$, $\bar n_1$, and $\bar n_2$ by
\begin{align}\label{barred ns}
	&\bar m = 
		\begin{cases}
		m & \text{if $m$ is odd,} \\
		2m & \text{if $m$ is even,}
		\end{cases}
	&&\bar n_j = 
		\begin{cases}
		n_j & \text{if $n_j$ is odd,} \\
		2n_j & \text{if $n_j$ is even.}
		\end{cases}
\end{align} 

Let $\HH^{m}$, $\HH^{n_1}$, and $\HH^{n_2}$ be Hilbert spaces with 
bases labelled by the elements in the rings $\ints_{m}$, $\ints_{n_1}$, and $\ints_{n_2}$, respectively.
The assignment $\ket{u}\to\ket{u_1}\otimes\ket{u_2}$
defines an isometry from $\HH^{m}$ onto 
$\HH^{n_1}\otimes\HH^{n_2}$.
We use this isomorphism to identify $\HH^{m}$ with 
$\HH^{n_1}\otimes\HH^{n_2}$.
The displacement operators on $\HH^{m}$ then split into pairs of displacement operators:
\begin{equation}\label{splitting}
	D_{a,b}^{(m)}=D_{a,\kappa_2b}^{(n_1)}\otimes D_{a,\kappa_1b}^{(n_2)}.
\end{equation}
The integers $\kappa_1$ and $\kappa_2$ are the multiplicative inverses of $n_1$ and $n_2$ 
in arithmetic 
modulo $\bar n_2$ and $\bar n_1$, respectively. That is,
$\kappa_1n_1=1\md \bar n_2$ and $\kappa_2n_2 = 1\md \bar n_1$.
To verify \eqref{splitting}, we calculate the action of the left-hand side operator on $\ket{u}$
and the action of the right-hand side operators
on $\ket{u_1}$ and $\ket{u_2}$.
The outcome is
\begin{align}
	&D_{a,b}^{(m)}\ket{u}
		=\tau_{m}^{ab}\omega_{m}^{ub}\ket{u+a},\\
	&D_{a,\kappa_2 b}^{(n_1)}\ket{u_1}
		=\tau_{n_1}^{ab\kappa_2}\omega_{n_1}^{u_1\kappa_2b}\ket{u_1+a_1},\\
	&D_{a,b\kappa_1}^{(n_2)}\ket{u_2}
		=\tau_{n_2}^{ab\kappa_1}\omega_{n_2}^{u_2b\kappa_1}\ket{u_2+a_2}.
\end{align}
Since $\ket{u+a}=\ket{u_1+a_1}\otimes \ket{u_2+a_2}$, it suffices to prove that
\begin{align}
	&\tau_{m} = \tau_{n_1}^{\kappa_2}\tau_{n_2}^{\kappa_1},\label{sjusju}\\
	&\omega_{m}^{u}=\omega_{n_1}^{u_1\kappa_2}\omega_{n_2}^{u_2\kappa_1}\label{sjuatta}.
\end{align}

To show that \eqref{sjusju} holds, we first observe that $\bar m=\bar n_1\bar n_2$ and that $\bar n_1$ and $\bar n_2$ are relatively prime. For $j=1,2$ 
define 
\begin{equation}
	\nu_j=
	\begin{cases}
		\kappa_j & \text{if $n_j$ is odd,} \\
		\frac{1}{2n_j}(m+n_j)\kappa_j& \text{if $n_j$ is even.}
	\end{cases}
\end{equation}
The numbers 
$\nu_1$ and $\nu_2$ are the multiplicative inverses of 
$\bar n_1$ and $\bar n_2$ in arithmetic modulo $\bar n_2$ and $\bar n_1$, respectively, and $\nu_1\bar n_1 + \nu_2\bar n_2 = 1 \md \bar m$.
Now, if $n_1$ and $n_2$ are both odd, then 
\begin{equation}
\begin{split}
	\tau_{m}^{\nu_1\bar n_1 + \nu_2\bar n_2} 	
	& = (-1)^{\nu_1\bar n_1 + \nu_2\bar n_2} 
		(e^{\frac{\pi i}{m}})^{\nu_1\bar n_1 + \nu_2\bar n_2} \\
	& =  (-1)^{\kappa_1 + \kappa_2} 
		(e^{\frac{\pi i}{n_2}})^{\kappa_1}
		(e^{\frac{\pi i}{n_1}})^{\kappa_2} \\
	& = \tau_{n_1}^{\kappa_2}\tau_{n_2}^{\kappa_1},
\end{split}
\end{equation}
and if one of $n_1$ or $n_2$ is even, e.g., if $n_1$ is even and $n_2$ is odd, then 
\begin{equation}
\begin{split}
	\tau_{m}^{\nu_1\bar n_1 + \nu_2\bar n_2} 	
	& = (-1)^{\nu_1\bar n_1 + \nu_2\bar n_2} 
		(e^{\frac{\pi i}{m}})^{\nu_1\bar n_1 + \nu_2\bar n_2} \\
	& = (-1)^{\kappa_2} 
		(e^{\frac{\pi i}{m}})^{2\nu_1n_1}
		(e^{\frac{\pi i}{n_1}})^{\kappa_2} \\
	& = \tau_{n_1}^{\kappa_2}(e^{\frac{\pi i}{n_2}})^{(n_2+1)\kappa_1} \\
	& = \tau_{n_1}^{\kappa_2}\tau_{n_2}^{\kappa_1}.
\end{split}
\end{equation}
This proves \eqref{sjusju}.
To prove \eqref{sjuatta}, we use the result in \eqref{sjusju} 
and calculate
\begin{equation}
	\omega_{m}^{u}
	=\tau_{m}^{2u}
	= \tau_{n_1}^{2u\kappa_2}\tau_{n_2}^{2u\kappa_1}
	= \omega_{n_1}^{u\kappa_2}\omega_{n_2}^{u\kappa_1}
	=\omega_{n_1}^{u_1\kappa_2}\omega_{n_2}^{u_2\kappa_1}.
\end{equation}
The last identity follows from $u\kappa_2=u_1\kappa_2\md n_1$ and 
$u\kappa_1=u_2\kappa_1\md n_2$.

\section{Expansions of $\Pi_1$ and $\Pi_2$} \label{expansion}
In this appendix we derive the expansion \eqref{utv1}
of $\Pi_1$ in the displacement operator basis. 
(The derivation of the expansion of $\Pi_2$ is similar so we omit it.)
Starting from Eq.\ \eqref{first projection},
\begin{equation}
\begin{split}
	\Pi_1
	&= \frac{d-1}{2d}\sum_{a,b=0}^{d-1} 
		D^{(n)}_{(d-2)a,(d-2)b} \ketbra{\psi_{0,0}}{\psi_{0,0}}D^{(n)}_{(2-d)a,(2-d)b} \\
	&= \frac{d-1}{2dn} \sum_{k,l=0}^{n^2-1} \sum_{a,b=0}^{d-1} 
		\bra{\psi_{0,0}}D_{k,l}^{(n)}\ket{\psi_{0,0}}
        D^{(n)}_{(d-2)a,(d-2)b} D^{(n)}_{-k,-l} D^{(n)}_{(2-d)a,(2-d)b} \\
	&= \frac{d-1}{2dn} \sum_{k,l=0}^{n^2-1} \sum_{a,b=0}^{d-1} 
        \bra{\psi_{0,0}}D_{k,l}^{(n)}\ket{\psi_{0,0}}
		 D^{(n)}_{-k,-l}  \omega_n^{(d-2)(lb-ka)} \\
	&=  \frac{d-1}{2dn} \sum_{k,l=0}^{n^2-1} \sum_{a,b=0}^{d-1} 
		 \bra{\psi_{0,0}}D_{k,l}^{(n)}\ket{\psi_{0,0}}
          D^{(n)}_{-k,-l}  \omega_d^{la-kb}.
\end{split}
\end{equation}
In the second identity, we have inserted the 
expansion of $\ketbra{\psi_{0,0}}{\psi_{0,0}}$ in the displacement operator basis.
In the third identity, we have used the commutation rule \eqref{commutation rule}.
Using that, for integer $m$, 
\begin{equation}\label{algebra}
	\sum_{a=0}^{d-1} \omega_d^{ma}=d\delta_{0,m}^{(d)},
\end{equation}
we can proceed and write
\begin{equation}
\begin{split}
	\Pi_1
    &=  \frac{d(d-1)}{2n} \sum_{k,l=0}^{n^2-1} 
		 \bra{\psi_{0,0}}D_{k,l}^{(n)}\ket{\psi_{0,0}}
          D^{(n)}_{-k,-l}  \delta_{0,l}^{(d)} \delta_{0,k}^{(d)} \\
    &=  \frac{d(d-1)}{2n} \sum_{a,b=0}^{d-3} 
		 \bra{\psi_{0,0}}D_{da,db}^{(n)}\ket{\psi_{0,0}} 
          D^{(n)}_{-da,-db}.
\end{split}
\end{equation}
This is the expansion in Eq.\ \eqref{utv1}.

\section{Parity operators}\label{Uniqueness of parity operators}
In this appendix, we show that the Clifford group contains only two parity operators,
namely $\pm P^{(n)}$ defined in Eq.\ \eqref{that guy}. To this end, let $P$ be any parity operator.
In \cite{Ap2005} it is shown that $P$, being a member of the Clifford group,
can be decomposed as $P=e^{i\theta}D_{k,l}^{(n)} V_F$.
Here, $F$ is a matrix in $\SL(2,\ints_{\bar n})$ and $V_F$ is the representation of $F$
defined in Section \ref{clifford}.

Suppose that
\begin{equation}
	F=\begin{pmatrix} \alpha & \beta \\ \gamma & \delta \end{pmatrix}.
\end{equation} 
Since $P$ is Hermitian, being an involution and a unitary, 
\begin{equation}\label{an X calculation}
	\1 
	= D_{k,l}^{(n)} V_F D_{1,0}^{(n)} V_F^\dagger D_{-k,-l}^{(n)} D_{1,0}^{(n)} 
	= \omega_n^{-l} D_{k,l}^{(n)} D_{\alpha,\gamma}^{(n)} D_{1,0}^{(n)} D_{-k,-l}^{(n)} 
	= \tau_n^{\gamma-2l} D_{k,l}^{(n)} D_{\alpha+1,\gamma}^{(n)} D_{-k,-l}^{(n)}.
\end{equation}
The second identity follows from \eqref{prop of F} and \eqref{commutation rule} and the third follows from \eqref{merging rule}.
Similarly, 
\begin{equation}\label{a Z calculation}
	\1 
	= D_{k,l}^{(n)} V_F D_{0,1}^{(n)} V_F^\dagger D_{-k,-l}^{(n)} D_{0,1}^{(n)} 
	= \omega_n^{k} D_{k,l}^{(n)} D_{\beta,\delta}^{(n)} D_{1,0}^{(n)} D_{-k,-l}^{(n)} 
	= \tau_n^{2k-\beta} D_{k,l}^{(n)} D_{\beta,\delta+1}^{(n)} D_{-k,-l}^{(n)}.
\end{equation}
These two calculations, together with the requirement that $\alpha\delta-\beta\gamma = 1 \md \bar n$,
show that if $n$ is odd, then
\begin{equation}\label{nittinio}
	\alpha+1=\beta=\gamma=\delta+1=k=l=0 \md n,
\end{equation}
while if $n$ is even, the multiple possible combinations for the entries of $F$ and the indices $k$ and $l$ are
the ones displayed in Table \ref{table1}.
(If $n$ is even, there is more than one option for the displacement operator in the fourth and 
eighth cases.
But the different displacement operators differ only by a sign which can be included in the phase 
factor $e^{i\theta}$.)
\begin{table}[t]
\centering
\begin{tabular}{|c|c|c|c|c|c|c|}
	\hline
	$\alpha$ & $\beta$ & $\gamma$ & $\delta$ & $k$ & $l$ 		\\
	\hline
	$-1\md \bar n$ & $0\md \bar n$ & $0\md \bar n$ & $-1\md \bar n$	& $0\md n$ & $0\md n$ \\    
	$-1\md \bar n$ & $0\md \bar n$ & $n\md \bar n$ & $-1\md \bar n$ & $0\md n$ & $n/2\md n$ \\
	$-1\md \bar n$ & $n\md \bar n$ & $0\md \bar n$ & $-1\md \bar n$ & $n/2\md n$ & $0\md n$ \\
	$-1\md \bar n$ & $n\md \bar n$ & $n\md \bar n$ & $-1\md \bar n$ & $n/2\md n$ & $n/2\md n$ \\
	$n-1\md \bar n$ & $0\md \bar n$ & $0\md \bar n$ & $n-1\md \bar n$ & $0\md n$ & $0\md n$ \\
	$n-1\md \bar n$ & $0\md \bar n$ & $n\md \bar n$ & $n-1\md \bar n$ & $0\md n$ & $n/2\md n$ \\
	$n-1\md \bar n$ & $n\md \bar n$ & $0\md \bar n$ & $n-1\md \bar n$ & $n/2\md n$ & $0\md n$ \\
	$n-1\md \bar n$ & $n\md \bar n$ & $n\md \bar n$ & $n-1\md \bar n$ & $n/2\md n$ & $n/2\md n$ \\
\hline
\end{tabular}
\caption{The possible values for the entries of $F$ and the indices $k,l$ of the displacement operator in 
the decomposition of $P$ when $n$ is even.}
\label{table1}
\end{table}

First, assume that $n$ is odd. According to \eqref{nittinio}, $P=e^{i\theta}V_F$ where 
\begin{equation}
	F=\begin{pmatrix} -1 & 0 \\ 0 & -1 \end{pmatrix}
	=\begin{pmatrix} 0 & -1 \\ 1 & 0 \end{pmatrix}
	\begin{pmatrix} 0 & -1 \\ 1 & 0 \end{pmatrix}
	=F_1F_2.
\end{equation}
By Eqs.\ \eqref{tut}, \eqref{the rep}, and \eqref{algebra},
\begin{equation}
	V_F	= \frac{1}{n}\sum_{u,v=0}^{n-1}\sum_{r,s=0}^{n-1}\omega_n^{uv+rs}\ket{u}\braket{v}{r}\bra{s}
		= \frac{1}{n}\sum_{u,s=0}^{n-1}\left(\sum_{v=0}^{n-1}\omega_n^{v(u+s)}\right)\ket{u}\bra{s}
		= P^{(n)},
\end{equation}
and the assumption $P^2=\1$ then forces the phase factor $e^{i\theta}$ to be $\pm 1$.
We conclude that $P=\pm P^{(n)}$.

Next, assume that $n$ is even. Then, by Table \ref{table1}, there are eight cases to check.
One can show that in all cases, $P=\pm P^{(n)}$.
Since the arguments are similar in all cases, we will do only one of the calculations, say, when
\begin{align}
	& \alpha=\delta=-1 \md \bar{n},
	&& \beta=\gamma=n\md \bar{n},
	&& k=l=n/2\md n.
\end{align}
This is the fourth row in Table \ref{table1}.
The decomposition 
\begin{equation}
	F=\begin{pmatrix} -1 & n \\ n & -1 \end{pmatrix}
	=\begin{pmatrix} 0 & -1 \\ 1 & 0 \end{pmatrix}
	\begin{pmatrix} n & -1 \\ 1 & -n \end{pmatrix}
	=F_1F_2
\end{equation}
is a prime decomposition of $F$ and, hence, by \eqref{tut} and \eqref{the rep},
\begin{equation}
	V_F	= \frac{1}{n}\sum_{u,v=0}^{n-1}\sum_{r,s=0}^{n-1}\omega_n^{uv}\tau_n^{-(ns^2-2rs-nr^2)}\ket{u}\braket{v}{r}\bra{s} 
		= \frac{1}{n}\sum_{u,v,s=0}^{n-1}\tau_n^{n(v^2-s^2)}\omega_n^{v(u+s)}\ket{u}\bra{s}.
\end{equation}
Using that 
\begin{equation}
	\tau_n^{n(v^2-s^2)}=(-1)^{(v-s)}
\end{equation}
and
\begin{equation}
	\sum_{v=0}^{n-1}(-1)^v\omega_n^{v(u+s)}=n\delta^{(n)}_{u+s,n/2},
\end{equation}
we can further reduce the expression for $V_F$:
\begin{equation}
	V_F	= \frac{1}{n}\sum_{u,s=0}^{n-1}(-1)^{s}
				\left(\sum_{v=0}^{n-1}(-1)^v\omega_n^{v(u+s)}\right)\ket{u}\bra{s}
		= \sum_{s=0}^{n-1}(-1)^{s}\ket{n/2-s}\bra{s}.
\end{equation}
Also, the displacement operator in the decomposition of $P$ is
\begin{equation}
	D^{(n)}_{n/2,n/2} = \tau_n^{n^2/4}X^{n/2}Z^{n/2} = i^{n/2}X^{n/2}Z^{n/2}.
\end{equation}
If we post-compose $V_F$ by this displacement operator, we obtain
\begin{equation}
\begin{split}
	D^{(n)}_{n/2,n/2}V_F 
	&= i^{n/2}\sum_{s=0}^{n-1}(-1)^{s}X^{n/2}Z^{n/2}\ket{n/2-s}\bra{s} \\
	&= i^{n/2}\sum_{s=0}^{n-1}(-1)^{s}\omega_n^{n^2/4-sn/2}\ket{-s}\bra{s} \\
	&= (-i)^{n/2}P^{(n)}.
\end{split}
\end{equation}
Then, finally, for $P=e^{i\theta}D^{(n)}_{n/2,n/2}V_F$ to be an involution, the phase factor $e^{i\theta}$ must be such that $(-i)^{n/2}e^{i\theta}=\pm 1$.
This finishes the proof that there is essentially only one parity operator,
namely $P^{(n)}$, regardless of the parity of $n$.

\section{Partial trace and local displacement operators}\label{Part Tr}
In this appendix, we prove Eqs.\ \eqref{partr1} and \eqref{partr2}.

Let $D_{a,b}^{(n_1)}$ and $D_{a,b}^{(n_2)}$ be the displacement operators 
corresponding to irreducible representations of $\WH(n_1)$ and $\WH(n_2)$
on an $n_1$-dimensional and an $n_2$-dimensional Hilbert space, respectively.
Then, for any operator $A$ on the composite Hilbert space,
\begin{align}
	\1_{n_1}\otimes\tr_{n_1}\!A
	&= \frac{1}{n_1} \sum_{a,b=0}^{n_1-1} 
			 (D^{(n_1)}_{-a,b}\otimes\1_{n_2}) A (D^{(n_1)}_{a,-b}\otimes\1_{n_2}), \label{pt1}\\
	\tr_{n_2}\!A\otimes\1_{n_2}
	&= \frac{1}{n_2} \sum_{a,b=0}^{n_2-1} 
			 (\1_{n_1} \otimes D^{(n_2)}_{a,b}) A (\1_{n_1}\otimes D^{(n_2)}_{-a,-b}). \label{pt2}
\end{align}

Before we prove Eq.\ \eqref{pt1} (the proof of \eqref{pt2} is similar)
we first prove that for any operator $B$ on the first factor,
\begin{equation}\label{tr}
	\frac{1}{n_1} \sum_{a,b=0}^{n_1-1} D^{(n_1)}_{-a,b} B D^{(n_1)}_{a,-b}=\tr B.
\end{equation}
We expand $B$ in the local displacement basis and use the commutation rule \eqref{commutation rule}
to conclude that
\begin{equation}
\begin{split}
	\frac{1}{n_1} \sum_{a,b=0}^{n_1-1} D^{(n_1)}_{-a,b} B D^{(n_1)}_{a,-b}
	&= \frac{1}{n_1^2} \sum_{k,l=0}^{n_1-1} \sum_{a,b=0}^{n_1-1} \tr(D^{(n_1)}_{k,l}B) \omega_{n_1}^{-bk-al} D^{(n_1)}_{-k,-l}.
\end{split}
\end{equation}
Equation \eqref{tr} now follows from the geometric sum \eqref{algebra}.

Next we prove Eq.\ \eqref{pt1}.
We begin by expanding $A$ in a product basis
\begin{equation}
	A = \sum_{k,k'=0}^{n_1-1}\sum_{l,l'=0}^{n_2-1} A_{k,k';l,l'} \ketbra{k}{k'}\otimes \ketbra{l}{l'}.
\end{equation}
If we then insert this expansion into the right-hand side of \eqref{pt1}
and apply \eqref{tr}, the right-hand side reduces to
\begin{equation}
\begin{split}
	\frac{1}{n_1} \sum_{a,b=0}^{n_1-1} \sum_{k,k'=0}^{n_1-1} \sum_{l,l'=0}^{n_2-1} 
	A_{k,k';l,l'} & D^{(n_1)}_{-a,b}  \ketbra{k}{k'} D^{(n_1)}_{a,-b} \otimes \ketbra{l}{l'}  \\
	&=  \sum_{k=0}^{n_1-1} \sum_{l,l'=0}^{n_2-1} A_{k,k;l,l'} \1_{n_1}\otimes\ketbra{l}{l'} 
	=\1_{n_1}\otimes\tr_{n_1}\!A.
\end{split}
\end{equation}
This proves \eqref{pt1}, from which Eq.\ \eqref{partr1} follows immediately. 
Equation \eqref{partr2} follows from \eqref{pt2}.

\section*{Acknowledgements}
The authors thank Ingemar Bengtsson for proposing the problem addressed in the current paper, 
for providing the representation in Appendix \ref{unusual}, for suggesting improvements to the text, 
and for numerous fruitful discussions. We also thank Marcus Appleby for sharing his notes on Chinese remaindering with us.

\end{document}